\documentclass[conference]{IEEEtran}
\usepackage[utf8]{inputenc}
\IEEEoverridecommandlockouts
\usepackage{cite}
\usepackage{amsmath,amssymb,amsfonts}
\usepackage{algorithmic}
\usepackage{graphicx}
\usepackage{textcomp}
\usepackage{comment}
\usepackage{xcolor}
\usepackage{multirow}

\newcommand{\doublefig}[2]{
\centering
\vspace{-3mm}
\subfloat[]{
\includegraphics[trim=0 22 0 40, clip, width=0.85\columnwidth]{#1}
}\\
\vspace{-3mm}
\subfloat[]{
\includegraphics[trim=0 22 0 40, clip, width=0.85\columnwidth]{#2}
}
\vspace{-1mm}
}
\usepackage{array}
\newcolumntype{L}[1]{>{\raggedright\let\newline\\\arraybackslash\hspace{0pt}}m{#1}}
\newcolumntype{C}[1]{>{\centering\let\newline\\\arraybackslash\hspace{0pt}}m{#1}}
\newcolumntype{R}[1]{>{\raggedleft\let\newline\\\arraybackslash\hspace{0pt}}m{#1}}

\ifCLASSOPTIONcompsoc
 \usepackage[caption=false,font=normalsize,labelfont=sf,textfont=sf]{subfig}
\else
 \usepackage[caption=false,font=footnotesize]{subfig}
\fi

\begin{document}

\title{Performance Assessment of Linear Models of Hydropower Plants
\thanks{This research was supported by the European Union Horizon 2020 research and innovation program in the context of the Hydropower Extending Power System Flexibility project (XFLEX HYDRO, grant agreement No 857832).}
}
%1, DITEN, Via all’Opera Pia 11a, Genova, Italy 

\author{\IEEEauthorblockN{Stefano Cassano, Fabrizio Sossan}
\IEEEauthorblockA{PERSEE, Mines ParisTech - PSL \\
Sophia Antipolis, France \\
\{stefano.cassano, fabrizio.sossan\}@mines-paristech.fr
}
\and
\IEEEauthorblockN{Christian Landry, Christophe Nicolet}
\IEEEauthorblockA{Power Vision Engineering Sarl\\
St-Sulpice, Switzerland \\
\{christian.landry, christophe.nicolet\}@powervision-eng.ch
}
}

%%% 
\begin{comment}
% Python script for equally long emails ..

from math import * 

fill_space = lambda tot, x : ''.join([empty_str(ceil((tot-len(x))/2)), x, empty_str(floor((tot-len(x))/2))])
empty_str = lambda x : ''.join(['~' for i in range(x)])
n = 40;
emails = 'fabio.dagostino@unige.it,stefano.massucco@unige.it,mario.paolone@epfl.ch,paola.pongiglione@hitachi-powergrids.com,federico.silvestro@unige.it,fabrizio.sossan@mines-paristech.fr'
for email in emails.split(','):
    fill_space(n, email)

\end{comment}
%%

\vspace{-10mm}
\maketitle

\begin{abstract}
This paper discusses linearized models of hydropower plants (HPPs). First, it reviews state-of-the-art models and discusses their non-linearities, then it proposes suitable linearization strategies for the plant head, discharge, and turbine torque. It is shown that neglecting the dependency of the hydroacoustic resistance on the discharge leads to a linear formulation of the hydraulic circuits model. For the turbine, a numerical linearization based on a first-order Taylor expansion is proposed. Model performance is evaluated for a medium- and a low-head HPP with a Francis and Kaplan turbine, respectively. Perspective applications of these linear models are in the context of efficient model predictive control of HPPs based on convex optimization.
\end{abstract}


%\tableofcontents

\section{Introduction}
Hydropower plants (HPPs) are a key renewable generation asset, covering more than 10\% of the electricity needs in Europe \cite{EuroStat}. Meanwhile, the increasing proportion of stochastic renewable generation in the power grid causes increasing regulation duties for conventional generation assets, including HPPs. Excessive regulation duties are a concern for HPP operators because they lead to increased wear and tear, ultimately shortening service life and requiring expensive maintenance. The need to counteract these effects has been very recently recognized in funded research projects (e.g., \cite{noauthor_hydropower_nodate}) and addressed in recent technical literature. E.g., work \cite{Dreyer2019DigitalCF} has shown that medium-head HPPs providing ancillary services incur in larger penstock fatigue, and authors of \cite{9209857} proposed a method to reduce it. As an alternative to extending regulation duties of HPPs, the use of batteries was proposed in so-called hybrid HPPs to increment the regulation capacity, e.g., \cite{9160666}.

% Literature review, including penstock nicolet \cite{Dreyer2019DigitalCF}, penstock nostro\cite{9209857}, i finlandesi con la batterie \cite{9160666} ed altri - se rilevanti - c'e' qualcosa dall'ultimo PSCC che puoi considerare?} 

Conventional HPP regulation loops include the droop governors for primary frequency regulation, the speed changer for secondary frequency control, and the turbine governor. The governor parameters are typically tuned to deliver the design performance (e.g., response time and droop) while respecting the plant's static mechanical and power limits. These classical feedback control loops do not model dynamic mechanical loads explicitly, so they are unaware of possible wear and tear effects that excessive regulation causes. Modeling the mechanical stress is relevant not only for wear and tear but also to design stress-informed splitting policies for the control signal in plants with multiple controllable elements, like hybrid HPPs.

An alternative to classical regulation loops to develop informed control decisions is model predictive control (MPC), which uses models to formulate constraints explicitly, as for example done in \cite{sossan2016achieving} for battery systems using linear prediction models of the battery voltage. In this spirit, this paper proposes linear models of the HPP that can be implemented into an MPC problem to formulate suitable operational constraints of the plant. Two linear models are proposed: a guide vane-to-torque model (key to model the plant's power output) and a guide vane-to-head model, which is essential to characterize mechanical loads and fatigue. By virtue of their linearity, the models allow for a tractable formulation of the MPC problem through convex optimization. These models contribute to advancing the state-of-the-art because typical HPP models for control applications are non-linear transfer-function models (e.g., \cite{kundur2007power}).

The rest of this paper is organized as follows: Section II describes HPP models, Section III describes the proposed linearization procedures, Section IV the methods for the performance evaluation, Section V presents the results and Section VI draws the main conclusions.

%Several countries worldwide have set out ambitious energy transition targets to increase the deployment of renewable generation and decommission fossil-fuel and nuclear power plants at the end of their service life. 
%Today, conventional power plants are responsible for providing the majority of grid ancillary services. Their displacement in favor of non-dispatchable resources, like stochastic renewables, will require to identify new providers of ancillary services. This need has also been recognized at the level of the European Union, which has supported in the recent years extensive research efforts aimed at investigating and enhancing the flexibility for future power systems, see e.g. the research projects EU-SysFlex, OSMOSE, and XFLEX Hydro \cite{noauthor_pan-european_nodate, noauthor_optimal_nodate, noauthor_hydropower_nodate}. 

\section{Modelling hydropower plants}
From a modelling perspective, HPPs feature two main components: hydraulic circuits and turbine, as described next.
%In this Section, the state-of-the-art modelling strategy based on the electrical analogy approach will be applied to describe the behaviour of the most important components of low- and medium-head hydro power plants equipped with a Francis pump-turbine without surge tank. The electrical analogy approach lays on the fact that hydraulic components can be modelled as RLC electrical circuits in which the voltage and the current correspond to the piezometric head $H$ and discharge $Q$.  The components discussed in this paper are the penstock and the hydraulic turbines. 

\subsection{Hydraulic circuits}
\begin{figure}[h!]
    \centering
    \includegraphics[trim=40 10 40 10, clip, width=1\columnwidth]{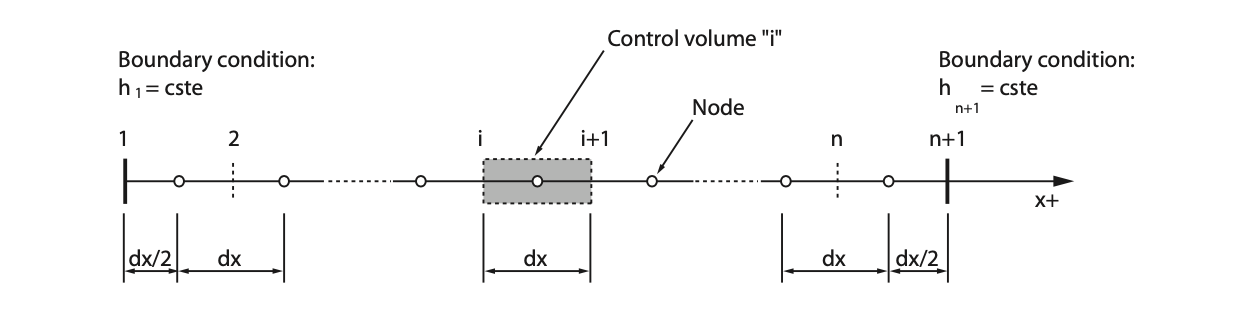}
    \caption{Spatial discretization of a pipe of length $L$ \cite{Nicolet:98534}.}
    \label{fig: Disc pipe}
\end{figure}

%A hydraulic circuit is a system comprising interconnected components that transport liquid. In an HPP, these are the conduits and the penstock.
%\todo{The difference between these two components lays on the complexity of the modelling of the dynamic behaviour of the water within the pipe. In this paper we take into the account the modelling of the penstock since it entails more detailed physics characterization. The conduits model can be directly derived from the former model by considering inelastic behaviour of the water. Moreover, other hydraulic components such as surge tanks, that are not discussed in the paper, can be modelled based on similar approach as the penstock.}
The hydraulic circuit of an HPP consists of the low-pressure tunnel, the penstock, and, in medium- and high-head plants, surge tanks. The penstock is the key element for dynamics because it is subject to the elastic behavior of the water. Its model is described next. The model of the surge tank can be derived by applying the same equivalent circuit principles described here and is not addressed in this paper for a reason of space.
The penstock is a pipe that guides water running from upstream to the hydraulic turbine.
The water's potential difference between the penstock's inlet and outlet is the source for the mechanical power. The penstock is not open to the air and is subject to water pressure. For this reason, water elasticity is accounted for in its model. Assuming that the penstock is significantly longer than larger, it can be modeled with a one-dimensional approach using partial differential equations (PDEs), e.g., \cite{Nicolet:98534}. PDEs are solved numerically by discretizing the penstock (Fig.~\ref{fig: Disc pipe}) into a finite number of elements, $n$, of length $dx=l/n$, where $l$ is the total length of the penstock.

This model can be conveniently visualized and solved in terms of its equivalent circuit model, where each element in Fig.~\ref{fig: Disc pipe} correspond to a (nonlinear) RLC circuit. The relationship between the discharge $Q_i$ of each penstock element $i$ and its head $h_i$ is:
\begin{subequations}
\begin{align}
& \dfrac{dQ_{i}}{dt}=-\dfrac{R(Q_i)}{L}\cdot Q_i-\dfrac{2}{L}\cdot h_{i+1/2}+\dfrac{2}{L}\cdot h_{i} \label{eqn:Simplified_41}\\
& \dfrac{dQ_{i+1}}{dt}=-\dfrac{R(Q_i)}{L}\cdot Q_{i+1}+\dfrac{2}{L}\cdot h_{i+1/2}-\dfrac{2}{L}\cdot h_{i+1} \label{eqn:Simplified_42}\\
& \dfrac{dh_{i+1/2}}{dt}=\dfrac{1}{C}\cdot (Q_{i}-Q_{i+1}). \label{eqn:Simplified_43}
\end{align}
where the circuit parameters are
\begin{align}
    R(Q_i) = \frac{\lambda\cdot |Q_i|\cdot dx}{2 g\cdot D\cdot A^{2}}, ~~
    L = \frac{dx}{g\cdot A}, ~~
    C = \frac{g\cdot A\cdot dx}{a^{2}},
    \label{eqn:RLC}
\end{align}
\end{subequations}
with $\lambda$ as the Darcy-Weisbach friction coefficient, $g$ acceleration of gravity, $A$ pipe cross-section, $D$ pipe diameter, and $a$ the wave speed in meters per second (m/s).

The equivalent circuit of a 1-element penstock model is shown in Fig.~\ref{fig:EEC}(b) and will be discussed later in combination with the turbine model. The number of penstock elements $n$ is chosen as a trade-off between computational complexity and modeling accuracy.

\subsection{Hydraulic turbines}
For dynamic power grid simulations, hydraulic turbines are typically modelled using the ``quasi-static'' approach, which assumes that the behavior of the hydraulic machines can be simulated as a succession of different steady-state conditions during the transition between different operating points \cite{Knapp2014CompleteC}. This approach, which preserves acceptable accuracy levels to model dynamic interactions with the power grid and is computationally tractable \cite{Nicolet:98534}, consists in using characteristic curves, typically determined experimentally, to link all operational variables of a turbine, namely its torque $T_t$, rotational speed $N$, head $H_t$, and flow $Q_t$. The characteristic curves are formulated in terms of the so-called unit variables:
\begin{align}
    N_{11}=\frac{N\cdot D_{n}}{\sqrt{H_t}}, ~~ Q_{11}=\frac{Q_t}{D_{n}^{2} \sqrt{H_t}}, ~~ T_{11}=\frac{T_t}{D_{n}^{2} H_t}
\end{align}
where $D_{n}$ is the diameter of the turbine. There are two characteristic curves, one for express the discharge factor $Q_{11}$, and the other for the torque factor $T_{11}$. Both are a function of the speed factor $N_{11}$ and the controllable inputs, which are, for Francis turbines, the guide vane $y$, and, for Kaplan turbines, the guide vane $y$ and the blade pitch $\beta$.

\subsubsection{Francis turbine}
\begin{figure}[h!]
    \centering
    \includegraphics[trim=0 20 0 20, width=0.95\columnwidth]{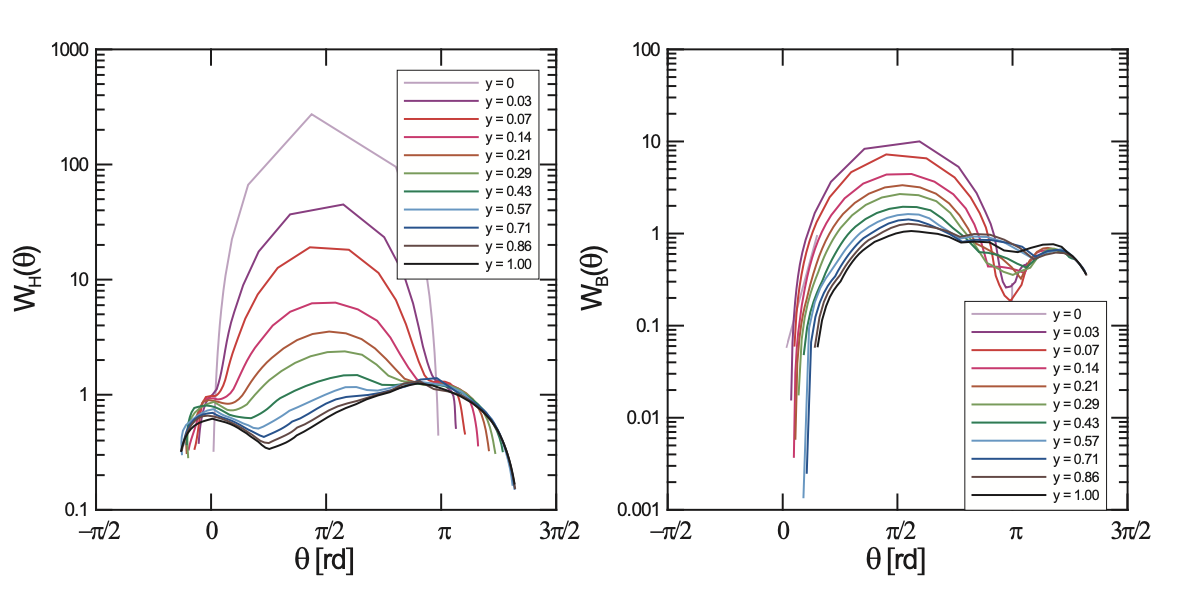}
    \caption{Polar representation of a Francis characteristic curves \cite{Nicolet:98534}.}
    \label{fig: Franc_char}
\end{figure}

As characteristic curves have typically an "S" shape, a change of variables is typically performed to avoid numerical issues \cite{10018994265}. This consists in defining a polar angle
\begin{subequations}\label{eq:characteristics}
\begin{align}
\theta(Q_t, N_t) = \arctan\left(\dfrac{Q_{11}/Q'_{11}}{N_{11}/N'_{11}}\right) = \arctan\left(\dfrac{Q_t}{N_t}\right), \label{eq:theta}
\end{align}
and two new functions of $\theta$ and guide vane $y$ defined as
\begin{align}
& W_H\left(\theta, y\right) = \dfrac{H_t/H'_t}{\left(Q_t/Q'_t\right)^2+\left(N/N'\right)^2} \label{eqn:W_H}\\
& W_B\left(\theta, y\right) = \dfrac{T_t/T_n}{\left(Q_t/Q'_t\right)^2+\left(N/N'\right)^2}, \label{eqn:W_B}
\end{align}
where $'$ quantities are values at the best efficiency point.
\end{subequations}
An example of these transformed characteristic curves are shown in Fig.~\ref{fig: Franc_char} and are the basis to derive the numerical first-order approximations for the linearized models.

%\todo{for a Francis turbine with specific speed $v=0.217$ \cite{Nicolet:98534}}

%A polar representation of the characteristic curves is commonly adopted to avoid numerical issues when interpolating the $Q_{11}(N_{11})$ curve due to its typical ``S" shape. 
%The polar representation, shown in Fig.~\ref{fig: Franc_char},  is defined in terms of polar angle $\theta = \arctan\left(\dfrac{Q_{11}/Q'_{11}}{N_{11}/N'_{11}}\right)$, where the subscript $n$ denotes nominal quantities, and two new variables

% \begin{align}
% \label{eqn:H_t}
% & H_t = H_n\cdot W_H\cdot \left[\left(\frac{Q_t}{Q_n}\right)^2+\left(\frac{N}{N_n}\right)^2\right]\\
% & T_t = T_n\cdot W_B\cdot \left[\left(\dfrac{Q}{Q_n}\right)^2+\left(\dfrac{N}{N_n}\right)^2\right], \label{eqn:T_t}
% \end{align}

\subsubsection{Kaplan turbine}\label{sec:kaplan}
Kaplan turbines feature a double control system comprising the guide vanes and mobile blades. Compared to the Francis turbine discussed above, their characteristic curves $W_H(\cdot), W_B(\cdot)$ are a function of the pitch angle $\beta$, too.

%Therefore, compared to the Francis turbine, they present an additional parameter which is the runner blade's pitch $\beta$. The polar representation of the Kaplan turbine is made by a family characteristics $W_H(\theta, y ,\beta)$ and $W_B(\theta, y ,\beta)$ for given values of $\beta$. Polar characteristics corresponding to runner blade angles different from the ones provided in the characteristic curves are obtained by linear interpolation between two given blade angles, $\beta_1$ and $\beta_2$.

\subsection{Complete plant model}
\begin{figure}
    \centering
    %\includegraphics[width=0.95\columnwidth]{Penstock_turbine scheme.png}\\
    %\footnotesize(a)
    %\hfill
    %\centering
    \includegraphics[width=0.85\columnwidth]{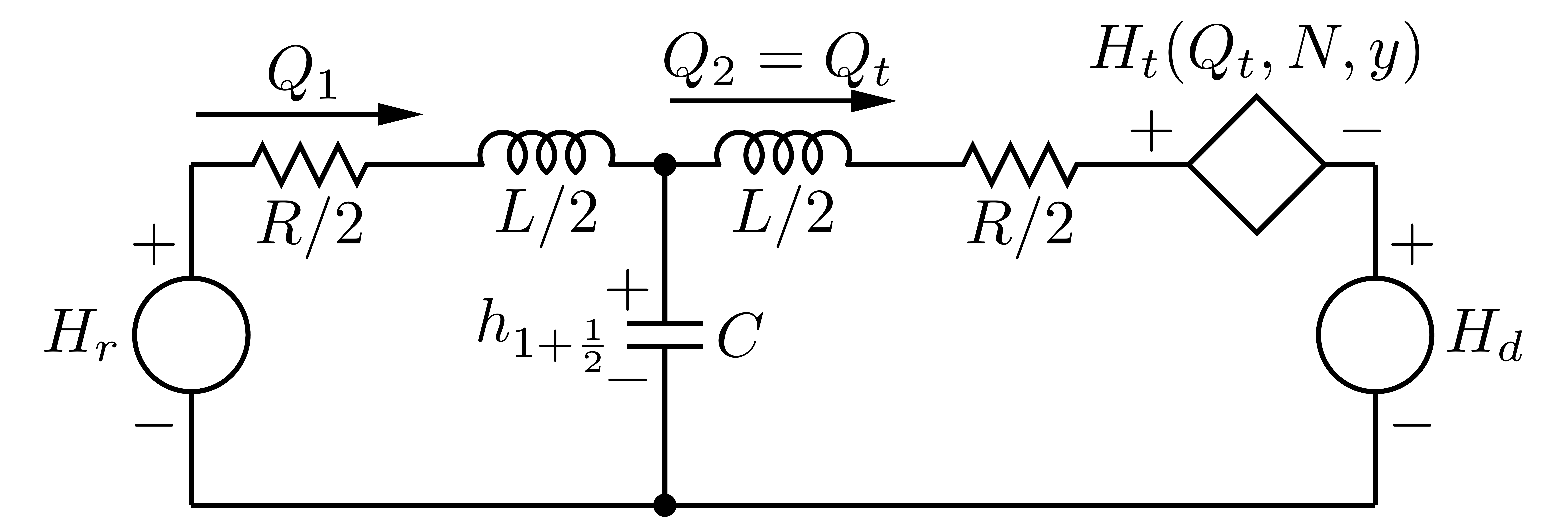}\\
    %\footnotesize(b)
    %\hfill
    \caption{Equivalent model of hydraulic circuits (with a 1-element penstock) and turbine.}\label{fig:EEC}
\end{figure}

The models of the hydraulic circuit and turbine can be combined in an equivalent circuit model \cite{hydraulic}, as shown in Fig.~\ref{fig:EEC}(b), where the RLC circuit refers to the (1-element) penstock and the variable voltage source $H_t$ to the turbine\footnote{In equivalent circuit models, voltages are analogous to pressures (or heads), and electric currents to the water flow.}, modelled with \eqref{eqn:W_H}. The inertia of the water and the no-discharge condition at guide vane full closure can be modelled with an equivalent inductance and a resistance in series to the turbine model, respectively \cite{bolleter1992hydraulic}  - not shown in Fig. \ref{fig:EEC} for a reason of space. It is convenient to write the equivalent circuit model in its state-space form to visualize all the involved quantities. The (augmented) state-space model also includes the rotational speed of the machine from Newton's second law for rotation. From the circuit with the 1-element penstock and Francis turbine of Fig. \ref{fig:EEC}(b), the state vector is:
\begin{subequations}\label{eq:statespace}
\begin{align}
x &=\begin{bmatrix} Q_1 & Q_t & h_{1+1/2} \;\; \omega\end{bmatrix}^\top
\end{align}
where $Q_1$ is the water flow in the penstock's first element, $Q_t$ is the turbine discharge, and $\omega$ the turbine angular velocity. The input vector is:
\begin{align}
u(Q_t, N, y) &= \begin{bmatrix} H_r \\ H_t(Q_t, N, y)-H_d \\ T_t(Q_t, N, y)-T_{el} \end{bmatrix} \label{eq:input}
\end{align}
where $H_r$ is the reservoir head, $H_d$ the downstream head, $T_{el}$ the electrical torque of the generator, $H_t$ and $T_{el}$ the turbine head and hydraulic torque from the characteristic curves in \eqref{eq:characteristics}, and $^\top$ denotes transpose. They both depend on the state components $Q_t$ and $N_t$, and guide vane opening $y$. The (nonlinear) state-space model is:
\begin{align}
\dot{x} = A(Q_i)x + B u(Q_t, N, y) \label{eq:ssd:1}.
\end{align}
The state and input transformation matrices are:
\begin{align}
 A(x) =
 \begin{bmatrix}
  \frac{R\left(Q_1\right)}{L} &  0 &  -\frac{2}{L} & 0\\
   0 & \frac{R\left(Q_t\right)}{L} & \frac{2}{L} & 0 \\
   \frac{1}{C} & -\frac{1}{C} & 0 & 0 \\
   0 & 0 & 0 & 0 \\
 \end{bmatrix} ,
 B = 
 \begin{bmatrix}
  \frac{2}{L} & 0 & 0\\
  0 & -\frac{2}{L} & 0\\
  0 & 0 & 0\\
  0 & 0 & \frac{1}{J}
 \end{bmatrix} \label{eq:systeminputmatrices}
% & C =
%  \begin{bmatrix}
% 1 \;\; 1 \;\; 1
% \end{bmatrix},
% D =
% \begin{bmatrix}
% R_s \;\; E
% \end{bmatrix}  \label{eq:voltage:statespaceCDG}
\end{align}
where $J$ is the turbine inertia. $A$ depends on the flow in the penstock and turbine discharge because of the dependency of the hydroacoustic resistance \eqref{eqn:RLC} on the flow.
\end{subequations}

When the penstock is modelled with $n$ elements, the state vector has $(2n+2)$ elements (i.e., $n+1$ discharges, $n$ heads, and 1 rotational speed). The model for the Kaplan turbine includes the dependency of $H_t$ and $T_t$ on blade angle $\beta$.

\section{Linear Models}
The state-space model in \eqref{eq:statespace} is non-linear in the state and in the controllable input, namely $y$ for the Francis turbine, and $y$ and $\beta$ for the Kaplan. We discuss in this section suitable methods and approximations to derive linearized models, whose performance are then investigated in the results section. The first non-linearity is the dependency between the hydroacoustic resistance $R$ and the discharge $Q$ in \eqref{eqn:RLC}. By assuming small variations of the operating point (and thus of the discharge), $R$ can be approximated as a constant value, finally leading to a linear and time-invariant formulation of the penstock model. The second non-linearity is the turbine's characteristic curves. As turbine models are derived from experimental measurements, a closed-form linearization is not possible, thus we proceed with a numerical linearization of the characteristic curves based on a first-order Taylor expansion, as discussed next.

\subsection{Linear model of a Francis turbine}
A relation between the head, denoted by $H_{t}(Q_t,N,y)$, and the numerical characteristic curve $W_h(\cdot)$ is obtained by inverting \eqref{eqn:W_H}. The procedure for the torque, not illustrated here, is analogue but considering \eqref{eqn:W_B}.

The first-order Taylor expansion of $H_{t}(Q_t,N,y)$, denoted by $\widetilde{H}_{t}(\cdot)$, around an operating point with discharge $Q_{t_0}$, rotational speed $N_0$, and guidevane opening $y_0$ reads as:
\begin{align}
\begin{aligned}
\widetilde{H}_{t}(Q_t,N,y) &\approx H_{t}(Q_{t_0}, N_0, y_0) + d^H_Q \cdot (Q_t-Q_{t_0}) + \\ & + d^H_N \cdot (N-N_0) + d^H_y \cdot (y-y_0). 
\end{aligned}\label{eq:linearhead}
\end{align}
where $d^H_Q, d^H_Q, d^H_N $ are partial derivatives of $H_{t}(Q_t,N,y)$ calculated in the operating point. They are computed by differentiating numerically as:
%are defined as $c_{(h,t)_Q}, c_{(h,t)_N}$ and $c_{(h,t)_y}$, where the lower case ``h" and ``t" stands for the head and torque case.
%The c- coefficient can be founded numerically. The next example shows the methodology for the assessment of the partial derivatives of $H_t$, the same approach is used for the torque.
\begin{subequations}
\begin{align}
    & d^H_Q := \dfrac{\partial H_t}{\partial Q_t}\bigg|_{Q_{t_0}} = \dfrac{H_t(Q_{t_0}+\epsilon, \cdot)-H_t(Q_{t_0}-\epsilon, \cdot)}{2\cdot \epsilon},\\
    & d^H_N := \dfrac{\partial H_t}{\partial N}\bigg|_{N_0} = \dfrac{H_t(N_0+\epsilon, \cdot)-H_t(N_0-\epsilon, \cdot)}{2\cdot \epsilon},\\
    & d^H_y := \dfrac{\partial H_t}{\partial y}\bigg|_{y_0} = \dfrac{H_t(y_0+\epsilon, \cdot)-H_t(y_0-\epsilon, \cdot)}{2\cdot \epsilon},
\end{align}
\end{subequations}
where $\epsilon$ is an arbitrary (small) parameter and $(\cdot)$ denote the remaining function arguments, kept constant.

\subsection{Linearized state-space model}
The state-dependant matrix $A(x)$ in \eqref{eq:systeminputmatrices} is calculated for the operating point $Q_{1_0}$ and $Q_{t_0}$, thus resulting in a linear and time-invariant transformation of the state.
The next step is expanding the linear models for the turbine head (and torque) in the input vector $u(\cdot)$ of \eqref{eq:input}. The turbine head appears in the second element of the input vector $u$; by using \eqref{eq:linearhead}, it can be re-written as:
\begin{subequations} \label{eqn:u2}
\begin{align}
  H_{t}(Q_t, N, y) \approx \begin{bmatrix}d^H_Q & d^H_N \end{bmatrix} M x + d^H_y y + c_H 
\end{align}
%\begin{align}
%\begin{aligned}
%     & T_{t}-T_{el} = u_{3} = [d^T_Q \hspace{0.1cm} d^T_N]\cdot M_{T}\cdot x + d^T_y\cdot y + \\ & (T_{t_{0}}-d^T_Q\cdot q_{0}-d^T_N\cdot n_{0}-d^T_y\cdot y_{0})-T_{el} \label{eqn:u3}
%     \end{aligned}   
%\end{align}
where $M$ is a $2 \times(2\cdot n+2)$ matrix such that $Mx = \begin{bmatrix} Q_t & N \end{bmatrix}^\top$, and $c_H$ collects all the known terms of the expression
\begin{align}
  c_H = H_{t}(Q_{t_0}, N_0, y_0) - d^H_Q Q_{t_0}-d^H_N N_{0}-d^H_y y_{0} 
\end{align}
\end{subequations}
Similarly, the turbine torque, appearing in the third term of the input vector $u(\cdot)$ in \eqref{eq:input}, can be written as:
\begin{subequations}\label{eqn:u3}
\begin{align}
& T_{t}(Q_t, N, y) \approx \begin{bmatrix}d^T_Q & d^T_N \end{bmatrix} M x + d^T_y  y + c_T \\
& c_{T} = T_{t}(Q_{t_0}, N_0, y_0) - d^T_Q Q_{t_0} - d^T_N N_0 - d^T_y y_0.
\end{align}
\end{subequations}

By replacing \eqref{eqn:u2} and \eqref{eqn:u3} in \eqref{eq:input}, the state-space in \eqref{eq:ssd:1} can be written as:
\begin{align}\label{eq:semifinal}
\begin{aligned}
   \dot{x} &= A x + B_{1} H_r + \\
   &+ B_{2}\cdot\left(\begin{bmatrix}d^H_Q & d^H_N \end{bmatrix} M x + d^H_y y + c_{H} - H_d\right) + \\
   &+ B_{3}\cdot\left(\begin{bmatrix}d^T_Q & d^T_N \end{bmatrix} M x + d^T_y y + c_{T} - T_{el}\right).% = \\
   %& = \left(A + B_{2} \begin{bmatrix}d^H_Q & d^H_N \end{bmatrix}  M + B_{3} \begin{bmatrix}d^T_Q & d^T_N \end{bmatrix} M \right) x \\ & + \Big[B_{1} \hspace{0.2cm} (B_{2}\cdot d^H_y+B_{3}\cdot d^T_y) \hspace{0.2cm} B_{2} \hspace{0.2cm} B_{3} \Big]\cdot \begin{bmatrix} u_1 \; y\; c_H \; c_T \end{bmatrix}^\top
   \end{aligned}
\end{align}
where $B_1, B_2$ and $B_3$ are respectively the first, second and third columns of matrix $B$ in \eqref{eq:systeminputmatrices}. Eq.~\eqref{eq:semifinal} can be now written as the following linear state-space
\begin{subequations}
\begin{align}
     \dot{x} = \widetilde{A} x + \widetilde{B} \widetilde{u}  \label{eqn:linss}
\end{align}
where:
\begin{align}
& \widetilde{u} = \begin{bmatrix} H_r & y & (c_H - H_d) & (c_T - T_{el}) \end{bmatrix}^\top \label{eq:input}\\
& \widetilde{A} = A + B_{2} \begin{bmatrix}d^H_Q & d^H_N \end{bmatrix} M + B_3 \begin{bmatrix}d^T_Q & d^T_N \end{bmatrix} M \\
& \widetilde{B} = 
\begin{bmatrix}
B_{1} & (B_{2}\cdot d^H_y+B_{3}\cdot d^T_y) & B_{2} & B_{3}
\end{bmatrix}. \label{eq:systeminputmatrices}
\end{align}
\end{subequations}
The state evolution in \eqref{eqn:linss} is now a linear function of the state and the controllable input, namely the guide vane $y$. The input vector \eqref{eq:input} contains, in addition to $y$, the reservoir head, the downstream head, electrical torque (these three are input parameters), and constant coefficients $c_H$ and $c_T$ that depend on the linearization.

% \begin{align}
% & A_{lin} = \Big(A + B_{2}\cdot[c_{q} \hspace{0.2cm} c_{n}]\cdot M_{H} + B_{3}\cdot[c_{qt} \hspace{0.2cm} c_{nt}]\cdot M_{T}\Big)\\
% & B_{lin} = \Big[B_{1} \hspace{0.2cm} (B_{2}\cdot c_{y}+B_{3}\cdot c_{yt}) \hspace{0.2cm} B_{2} \hspace{0.2cm} B_{3} \Big]
% \end{align}

\subsection{The case of Kaplan turbines}
As discussed in Section \ref{sec:kaplan}, Kaplan turbines can adjust the blade pitch, $\beta$, too. The linearization is performed similarly to the Francis turbine, with an additional partial derivative for $\beta$. The linear state-space system for the Kaplan, $\widetilde{u}', \widetilde{A}', \widetilde{B}'$, is
\begin{subequations}
\begin{align} \label{eqn:u_lin_kapl}
& \widetilde{u}' = \begin{bmatrix} H_r & y & \beta & (c'_H - H_d) & (c'_T - T_{el}) \end{bmatrix}^\top \\
& \widetilde{A}' = \widetilde{A} \\
& \widetilde{B}' = 
\begin{bmatrix}
B_{1} & (B_{2} d^H_y+B_{3} d^T_y) & (B_{2} d^H_\beta+B_{3} d^T_\beta) & B_{3}\end{bmatrix} \label{eqn:B_kaplan}
\end{align}
where the known terms contain also the linearazition point of the blade pitch $\beta_0$:
\begin{align}
c'_H = c_H - d^H_\beta \beta_{0} \;\; \text{and} \;\; c'_T = c_T - d^T_\beta \beta_{0}.
\end{align}
\end{subequations}

\section{Methods for Performance Evaluation}

\subsection{Case studies}
We consider two HPPs of different kind. The first is 87 MW medium-head plant with a Francis turbine. It has a 500 meters penstock and a net head (i.e., $H_r - H_d$) of 90 meters. The penstock is discretized with $n=20$ elements. The second HPP is 39 MW Kaplan low-head unit with a net head of 15m. The short penstock and spiral case is modelled with 8 components.

\subsection{Procedure to compute the estimation performance}
The estimation performance of the linear models is evaluated in time domain simulations by comparing their output against the non-linear models. The procedure to compute the estimation performance is the following:

{\bf(Step 1)} a linear model is computed for each given operating point. The operating point is specified by the guide vane opening, net head, and rotational speed. Nine different operational points are considered, given by varying the guide vane from 0.2 pu to 1 pu (0 pu and 1 pu represent respectively the all close and all open position), representing the typical operating range of a power plant, with increments of 0.1 pu. The net-head is assumed constant at its nominal value, and the rotational speed at 50 Hz to represent steady-state synchronous operations. For the low-head HPP with Kaplan turbine (that features two regulation mechanisms, guide vane and blade bitch), the pitch is chosen as a function of the guide vane according to the on-cam curve;

{\bf(Step 2)} each linear model is used to simulate operations for a stepwise change of the guide vane. We consider 20 different stepwise changes, from -0.5 pu to 0.5 pu, with increments of 0.025 pu. All the combinations resulting in unfeasible guide vane openings (e.g., guide vane 0.2 and stepwise change of 0.5) are excluded. In this way, a total of 41 x 9 (minus the unfeasible combinations) simulations are performed. For the Kaplan turbine, each guide vane deviation determines a deviation of blade angle according to the on-cam curve.

{\bf(Step 3)} for each simulation, the estimation error is calculated as the difference between the linear model and the ground-truth model.
    
We analyze estimates of the the turbine torque (relevant in the context of characterizing the mechanical and electric power of the plant) and the spatially averaged head in the penstock for the medium-head HPP, and the head at the turbine for the low-head plant (relevant to asses mechanical load levels, and fatigue, of HPPs).

%As explained and formalized next, both the steady-state and transient performance of the linearized model is assessed.

\subsection{Performance metrics and notation}
We formalize the notions explained in the former section with the objective of defining the metrics. 
Let $\Psi$ denote the set with all linearized models of the low-head (or medium-head) HPP, and $\psi \in \Psi$ a single linearized model. 
Let set $X$ denote all possible combinations of linearization points and deviations of guide vane performed in the experiments, where $\chi \in X$ is a single experiment;
$\widehat{y}_T(t, \psi, \chi)$ is a time series that contains the turbine torque of the time-domain simulation of linear model $\psi$ for experiment $\chi$; $y_T(t, \chi)$ is the ground-truth time series from the non-linear model for the same experiment. The torque error of linear model $\psi$ in experiment $\chi$ is:
\begin{align}
    e_T(t, \psi, \chi) = \frac{y_T(t, \chi) - \hat{y}_T(t, \psi, \chi)}{T_n}\label{eq:mae}
    %e_H(t, \psi, \chi) = \frac{y_H(t, \chi) - \hat{y}_H(t, \psi, \chi)}{H_\text{nom}},
\end{align}
where $T_n$ is the nominal torque.
The torque estimation performance is evaluated in terms of the mean absolute error (MAE) of the error $e$:
\begin{align}
\text{MAE} = \sum_{t=t_0}^{t=t_f}  \left| e(t, \psi, \chi) \right|
\end{align}
where the initial and ending time intervals $(t_0, t_f)$ are chosen to either capture transient or steady-state conditions. For transient conditions, $t_0$ corresponds to when the step-wise change is applied and $t_f=t_0 + 350~s$, where 350~s is determined by empirically by evaluating steady-state conditions ($\dot x \approx 0$). For steady-state conditions, $t_0$ and $t_f$ are  set to a fixed time interval that correspond to when the system is in steady-state. Head estimations are computed and characterized with the same procedure, scaling the error by the nominal head $H_n$.

%where $x(t, p)$ is the ground-truth reference value, computed with the non-linear model.

%The performance metrics discussed in the following are the mean absolute spatially-average error
%\begin{align}
%\text{MAE} = \sum_{t=t_0}^{t=t_f} \left| \frac{1}{P}\sum_{p=1}^{P}  e(\chi, t, p) \right|
%\end{align}
%where the term in the absolute value operator is the average head error in the penstock,

%By properly adjusting the start and end time intervals, we assess the estimation performance to a step change of the guide vane opening during both steady-state and transient conditions of the step response.

\section{Results}
\subsection{Medium-head plant with Francis turbine}
\begin{figure}[h!]
\doublefig{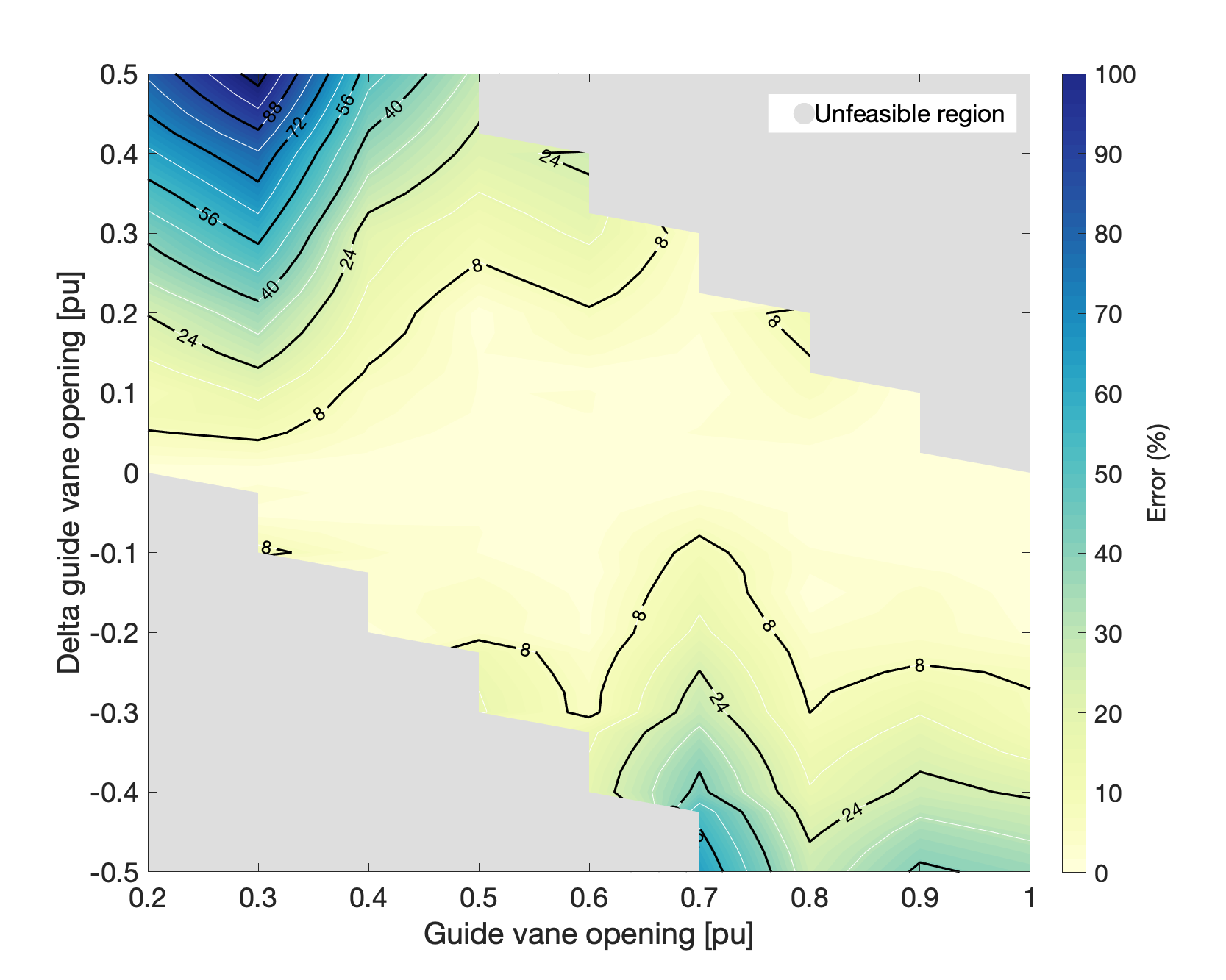}{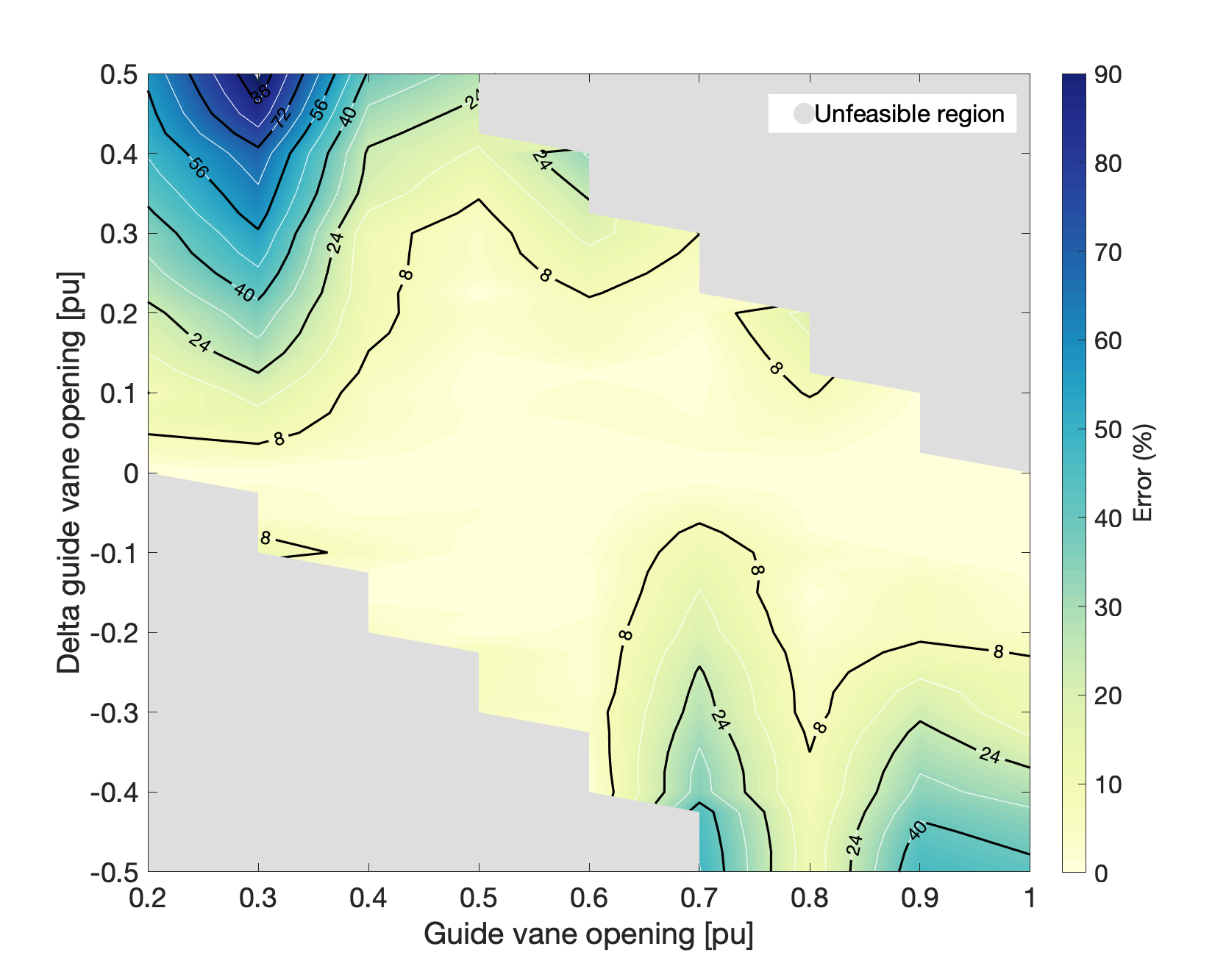}
\caption{Francis medium-head HPP, turbine torque: MAE of the linear estimations in transient (a) and steady-state (b) conditions.}\label{fig:francis_torque}
\end{figure}

\begin{figure}[h!]
\doublefig{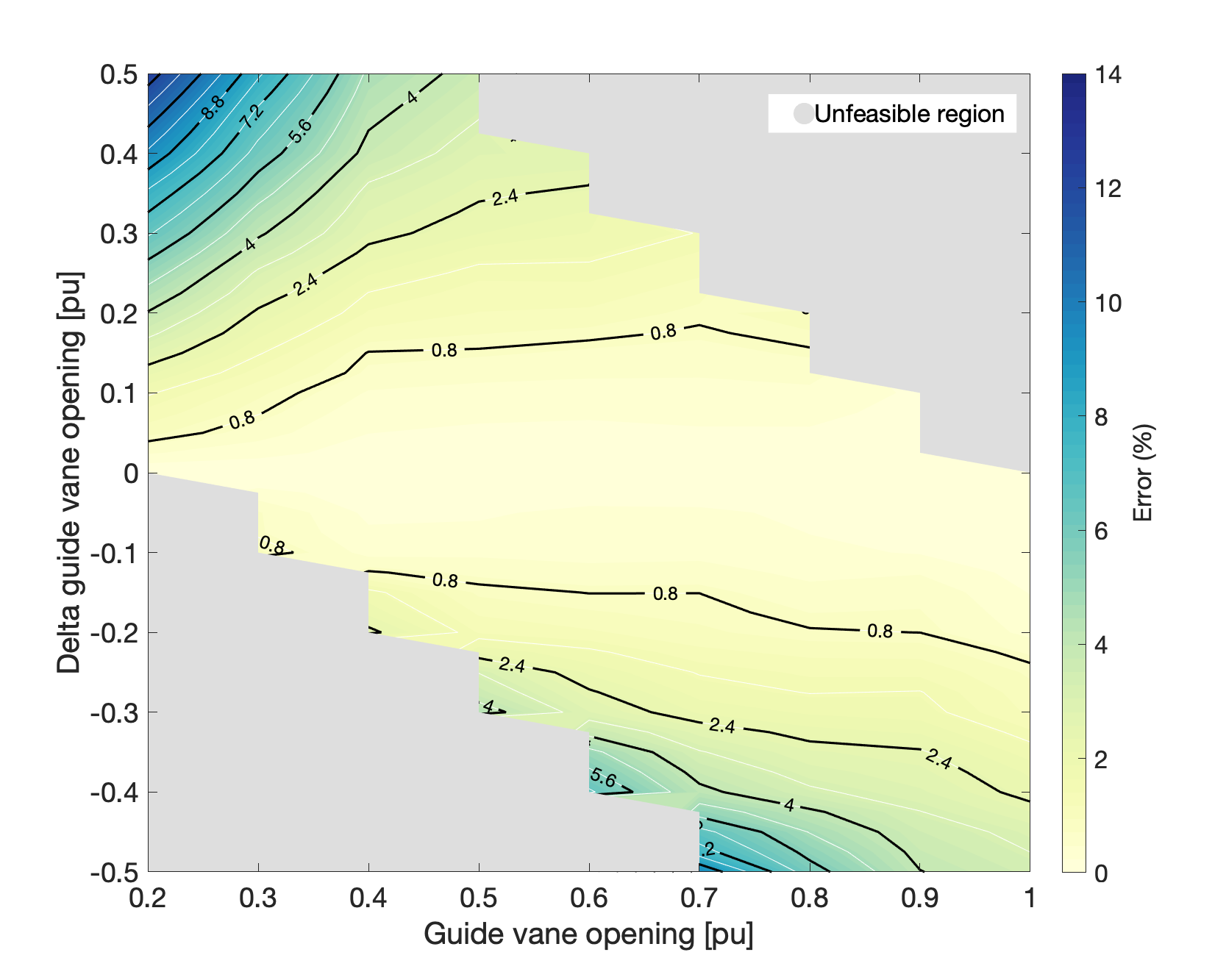}{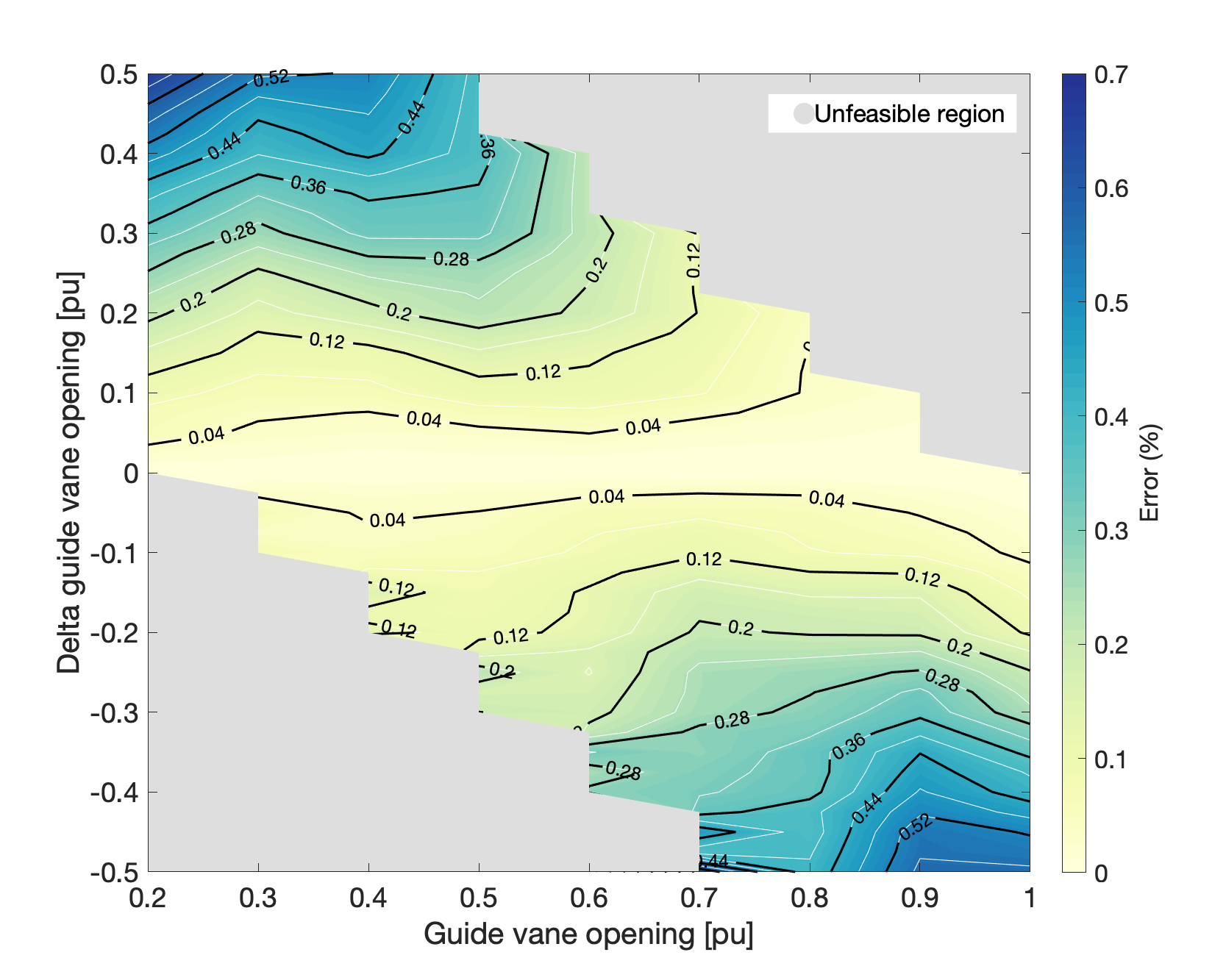}
\caption{Francis medium-head HPP, spatially averaged head in the penstock: MAE of the linear estimations in transient (a) and steady-state (b) conditions.}\label{fig:francis_head}
\end{figure}

Figures \ref{fig:francis_torque} and \ref{fig:francis_head} show the transient and steady-state MAE of the torque and head, respectively for the medium-head HPP with Kaplan. The main considerations that can be derived are discussed in the following findings.

{\bf Finding 1:} Linear estimates are more reliable for the head than for the torque. This is due to more prominent non-linearities in the torque model.

{\bf Finding 2:} For a given guide vane opening, estimation performance worsens with larger step-wise variations of the guide vane. This result is to be expected because the linear models are first-order approximations of the nonlinear models; thus, small deviations from the linearization point imply better local approximation.

{\bf Finding 3:} Linear head estimates are better in steady-state (with a maximum error of 1\%) than in transient conditions (maximum error: 10\%). For the torque, performance is similar in both cases. 

{\bf Finding 4:} For variations of the guide vane smaller than 0.1 pu, torque estimation errors are less than 10\% (except for guide vane 0.7 and less than 0.4), and head estimation errors less than 1\%.

In the light of Finding 4, it can be concluded that the use of linear models is justified in small-signal applications and where these error levels are acceptable; the advantage is handling computationally tractable models that can be implemented in, for example, (convex) optimization problems for optimal decision-making.

\subsection{Low-head plant with Kaplan turbine}
\begin{figure}[h!]
\doublefig{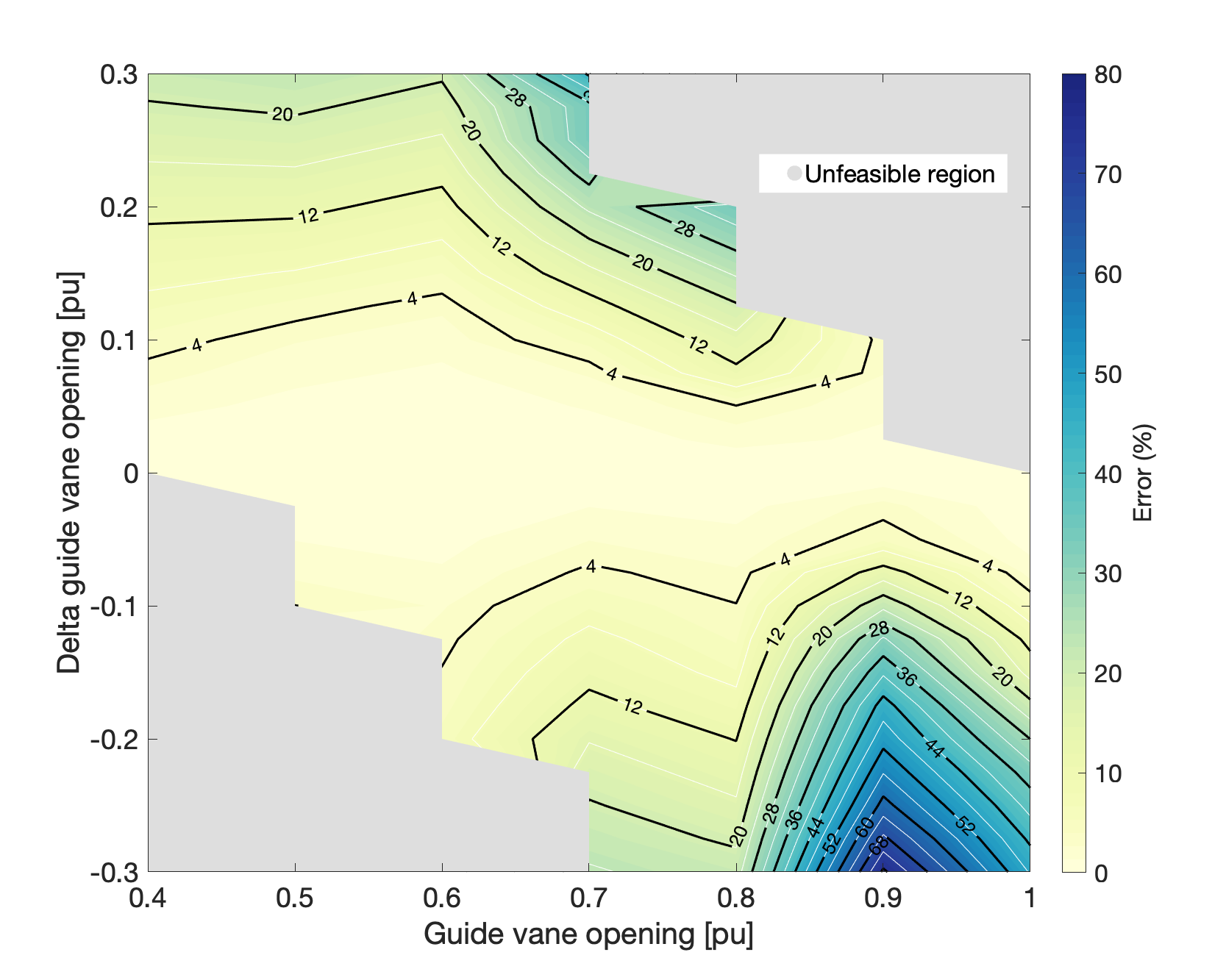}{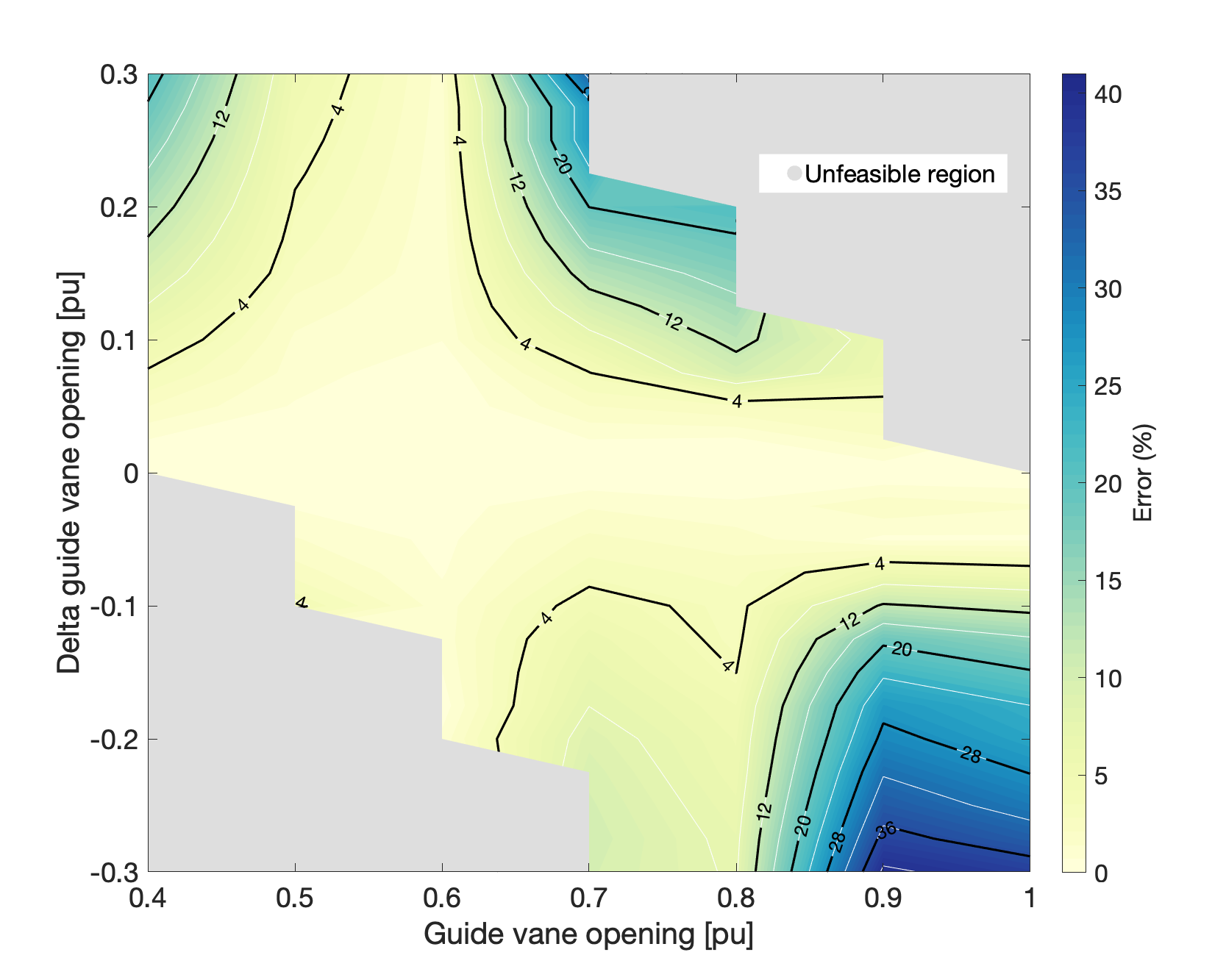}
\caption{Kaplan low-head HPP, turbine torque: MAE of the linear estimations in transient (a) and steady-state (b) conditions.}\label{fig:kaplan_torque}
\end{figure}

\begin{figure}[h!]
\doublefig{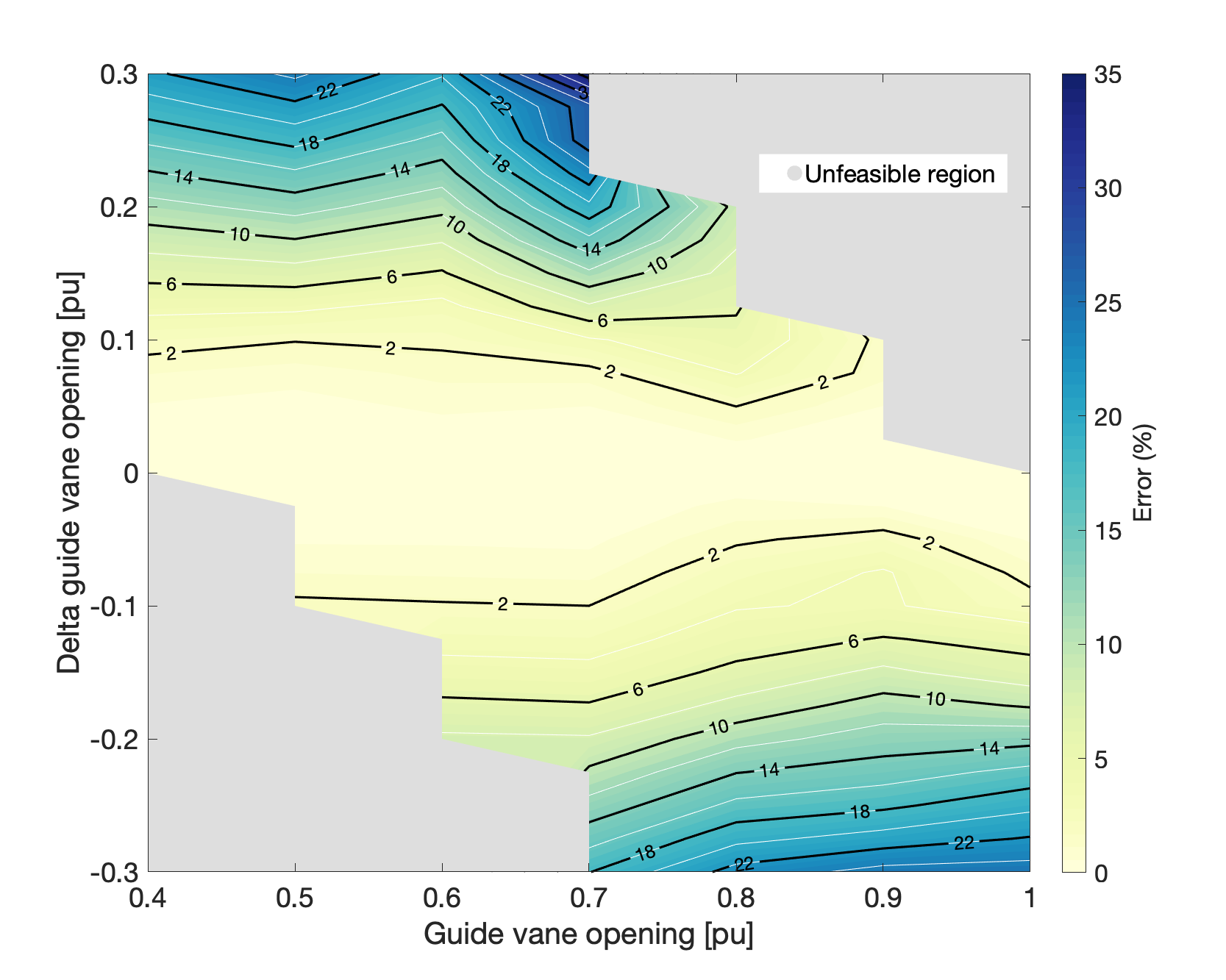}{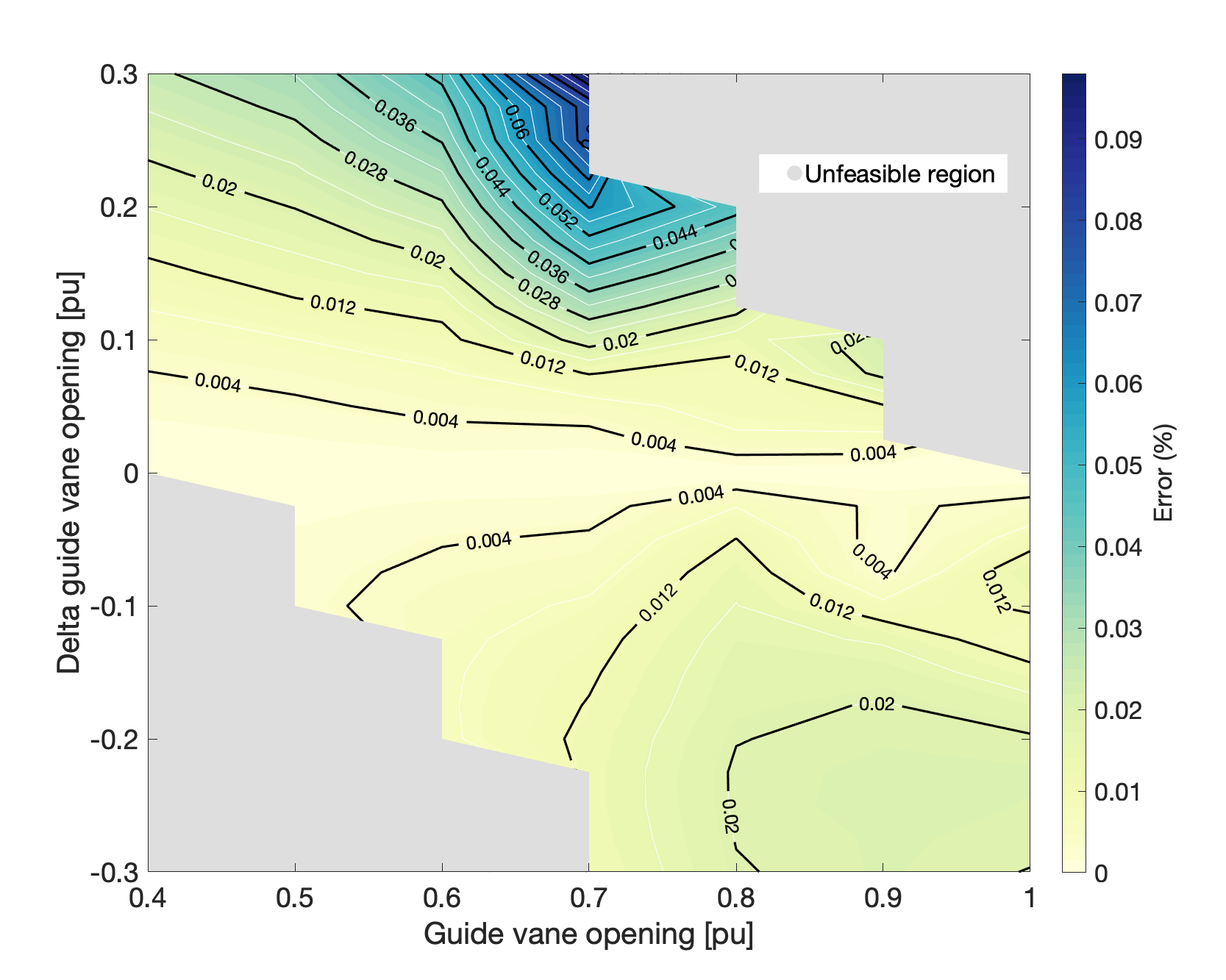}
\caption{Kaplan low-head HPP, head at the turbine inlet: MAE linear estimations in transient (a) and steady-state (b) conditions.}\label{fig:kaplan_head}
\end{figure}

Transient and steady-state performance of torque and head estimations is respectively shown in figures \ref{fig:kaplan_torque} and \ref{fig:kaplan_head}. Similar observations as for the medium-head HPP can be drawn. In particular:
(Finding 1) Linear estimations of the head are more accurate than for torque; (Finding 2) Smaller deviations of stepwise changes result in smaller estimations errors; (Finding 3) head estimates at steady state are significantly more reliable (errors less than 0.1\%) than during transients (errors less than 30\%). For the torque, there is no significant difference between the two cases; (Finding 4) for small signal variations (e.g., $\pm$0.1 pu of the guidavane), torque errors are approximately within a 10\% band, and head errors within 0.8\%, thus denoting relatively small errors of the linear models when used in small-signal applications.

\section{Conclusions}
This paper presented linearized models of hydropower plants and discussed their performance. The main sources of nonlinearities in both the hydraulic circuit and turbine models were illustrated, and linearization strategies were discussed. Estimation performance was investigated for both medium- and low-head HPPs with Francis and a Kaplan turbine, respectively. Results showed that i) linear estimates of the head are more reliable than for the torque; ii) for variations of the controllable input in a $\pm$0.1 pu range, the relative mean absolute error of the linear estimates are less than 10\% for the torque and less than 1\% for the head. In small signal applications where these error levels are considered acceptable, the linear models are a more tractable alternative to non-linear, opening to the development of efficient model predictive control based on convex optimization.

\bibliographystyle{IEEEtran}
\bibliography{biblio.bib}
\vfill

\end{document}